\documentclass[sigconf, screen]{acmart}
\AtBeginDocument{%
  \providecommand\BibTeX{{%
    \normalfont B\kern-0.5em{\scshape i\kern-0.25em b}\kern-0.8em\TeX}}}

\setcopyright{rightsretained}
\copyrightyear{2024}
\acmYear{2024}

\acmConference[arXiv]{arXiv: arXiv Preprint}{2024}{Atlanta, GA}
\acmBooktitle{arXiv: arXiv Preprint, 2024, Atlanta, GA}

\usepackage{subcaption}
\usepackage{enumitem}
\usepackage{rotating}
\usepackage{microtype}
\usepackage{bm}
\usepackage{soul}
\usepackage{wrapfig}

\usepackage[varqu]{zi4}
\usepackage{CJKutf8}
\usepackage{units}
\usepackage{booktabs}
\usepackage{colortbl}

\usepackage{amsmath}
\usepackage{mathtools}
\usepackage{mathtools}

\definecolor{redOV}{RGB}{255, 235, 238}
\definecolor{redI}{RGB}{255, 205, 210}
\definecolor{redII}{RGB}{239, 154, 154}
\definecolor{redIII}{RGB}{229, 115, 115}
\definecolor{redIV}{RGB}{239, 83, 80}
\definecolor{redV}{RGB}{244, 67, 54}
\definecolor{redVI}{RGB}{229, 57, 53}
\definecolor{redVII}{RGB}{211, 47, 47}
\definecolor{redVIII}{RGB}{198, 40, 40}
\definecolor{redIX}{RGB}{183, 28, 28}
\definecolor{redAI}{RGB}{255, 138, 128}
\definecolor{redAII}{RGB}{255, 82, 82}
\definecolor{redAIV}{RGB}{255, 23, 68}
\definecolor{redAVII}{RGB}{213, 0, 0}

\definecolor{pinkOV}{RGB}{252, 228, 236}
\definecolor{pinkI}{RGB}{248, 187, 208}
\definecolor{pinkII}{RGB}{244, 143, 177}
\definecolor{pinkIII}{RGB}{240, 98, 146}
\definecolor{pinkIV}{RGB}{236, 64, 122}
\definecolor{pinkV}{RGB}{233, 30, 99}
\definecolor{pinkVI}{RGB}{216, 27, 96}
\definecolor{pinkVII}{RGB}{194, 24, 91}
\definecolor{pinkVIII}{RGB}{173, 20, 87}
\definecolor{pinkIX}{RGB}{136, 14, 79}
\definecolor{pinkAI}{RGB}{255, 128, 171}
\definecolor{pinkAII}{RGB}{255, 64, 129}
\definecolor{pinkAIV}{RGB}{245, 0, 87}
\definecolor{pinkAVII}{RGB}{197, 17, 98}

\definecolor{purpleOV}{RGB}{243, 229, 245}
\definecolor{purpleI}{RGB}{225, 190, 231}
\definecolor{purpleII}{RGB}{206, 147, 216}
\definecolor{purpleIII}{RGB}{186, 104, 200}
\definecolor{purpleIV}{RGB}{171, 71, 188}
\definecolor{purpleV}{RGB}{156, 39, 176}
\definecolor{purpleVI}{RGB}{142, 36, 170}
\definecolor{purpleVII}{RGB}{123, 31, 162}
\definecolor{purpleVIII}{RGB}{106, 27, 154}
\definecolor{purpleIX}{RGB}{74, 20, 140}
\definecolor{purpleAI}{RGB}{234, 128, 252}
\definecolor{purpleAII}{RGB}{224, 64, 251}
\definecolor{purpleAIV}{RGB}{213, 0, 249}
\definecolor{purpleAVII}{RGB}{170, 0, 255}

\definecolor{deeppurpleOV}{RGB}{237, 231, 246}
\definecolor{deeppurpleI}{RGB}{209, 196, 233}
\definecolor{deeppurpleII}{RGB}{179, 157, 219}
\definecolor{deeppurpleIII}{RGB}{149, 117, 205}
\definecolor{deeppurpleIV}{RGB}{126, 87, 194}
\definecolor{deeppurpleV}{RGB}{103, 58, 183}
\definecolor{deeppurpleVI}{RGB}{94, 53, 177}
\definecolor{deeppurpleVII}{RGB}{81, 45, 168}
\definecolor{deeppurpleVIII}{RGB}{69, 39, 160}
\definecolor{deeppurpleIX}{RGB}{49, 27, 146}
\definecolor{deeppurpleAI}{RGB}{179, 136, 255}
\definecolor{deeppurpleAII}{RGB}{124, 77, 255}
\definecolor{deeppurpleAIV}{RGB}{101, 31, 255}
\definecolor{deeppurpleAVII}{RGB}{98, 0, 234}

\definecolor{indigoOV}{RGB}{232, 234, 246}
\definecolor{indigoI}{RGB}{197, 202, 233}
\definecolor{indigoII}{RGB}{159, 168, 218}
\definecolor{indigoIII}{RGB}{121, 134, 203}
\definecolor{indigoIV}{RGB}{92, 107, 192}
\definecolor{indigoV}{RGB}{63, 81, 181}
\definecolor{indigoVI}{RGB}{57, 73, 171}
\definecolor{indigoVII}{RGB}{48, 63, 159}
\definecolor{indigoVIII}{RGB}{40, 53, 147}
\definecolor{indigoIX}{RGB}{26, 35, 126}
\definecolor{indigoAI}{RGB}{140, 158, 255}
\definecolor{indigoAII}{RGB}{83, 109, 254}
\definecolor{indigoAIV}{RGB}{61, 90, 254}
\definecolor{indigoAVII}{RGB}{48, 79, 254}

\definecolor{blueOV}{RGB}{227, 242, 253}
\definecolor{blueI}{RGB}{187, 222, 251}
\definecolor{blueII}{RGB}{144, 202, 249}
\definecolor{blueIII}{RGB}{100, 181, 246}
\definecolor{blueIV}{RGB}{66, 165, 245}
\definecolor{blueV}{RGB}{33, 150, 243}
\definecolor{blueVI}{RGB}{30, 136, 229}
\definecolor{blueVII}{RGB}{25, 118, 210}
\definecolor{blueVIII}{RGB}{21, 101, 192}
\definecolor{blueIX}{RGB}{13, 71, 161}
\definecolor{blueAI}{RGB}{130, 177, 255}
\definecolor{blueAII}{RGB}{68, 138, 255}
\definecolor{blueAIV}{RGB}{41, 121, 255}
\definecolor{blueAVII}{RGB}{41, 98, 255}

\definecolor{lightblueOV}{RGB}{225, 245, 254}
\definecolor{lightblueI}{RGB}{179, 229, 252}
\definecolor{lightblueII}{RGB}{129, 212, 250}
\definecolor{lightblueIII}{RGB}{79, 195, 247}
\definecolor{lightblueIV}{RGB}{41, 182, 246}
\definecolor{lightblueV}{RGB}{3, 169, 244}
\definecolor{lightblueVI}{RGB}{3, 155, 229}
\definecolor{lightblueVII}{RGB}{2, 136, 209}
\definecolor{lightblueVIII}{RGB}{2, 119, 189}
\definecolor{lightblueIX}{RGB}{1, 87, 155}
\definecolor{lightblueAI}{RGB}{128, 216, 255}
\definecolor{lightblueAII}{RGB}{64, 196, 255}
\definecolor{lightblueAIV}{RGB}{0, 176, 255}
\definecolor{lightblueAVII}{RGB}{0, 145, 234}

\definecolor{cyanOV}{RGB}{224, 247, 250}
\definecolor{cyanI}{RGB}{178, 235, 242}
\definecolor{cyanII}{RGB}{128, 222, 234}
\definecolor{cyanIII}{RGB}{77, 208, 225}
\definecolor{cyanIV}{RGB}{38, 198, 218}
\definecolor{cyanV}{RGB}{0, 188, 212}
\definecolor{cyanVI}{RGB}{0, 172, 193}
\definecolor{cyanVII}{RGB}{0, 151, 167}
\definecolor{cyanVIII}{RGB}{0, 131, 143}
\definecolor{cyanIX}{RGB}{0, 96, 100}
\definecolor{cyanAI}{RGB}{132, 255, 255}
\definecolor{cyanAII}{RGB}{24, 255, 255}
\definecolor{cyanAIV}{RGB}{0, 229, 255}
\definecolor{cyanAVII}{RGB}{0, 184, 212}

\definecolor{tealOV}{RGB}{224, 242, 241}
\definecolor{tealI}{RGB}{178, 223, 219}
\definecolor{tealII}{RGB}{128, 203, 196}
\definecolor{tealIII}{RGB}{77, 182, 172}
\definecolor{tealIV}{RGB}{38, 166, 154}
\definecolor{tealV}{RGB}{0, 150, 136}
\definecolor{tealVI}{RGB}{0, 137, 123}
\definecolor{tealVII}{RGB}{0, 121, 107}
\definecolor{tealVIII}{RGB}{0, 105, 92}
\definecolor{tealIX}{RGB}{0, 77, 64}
\definecolor{tealAI}{RGB}{167, 255, 235}
\definecolor{tealAII}{RGB}{100, 255, 218}
\definecolor{tealAIV}{RGB}{29, 233, 182}
\definecolor{tealAVII}{RGB}{0, 191, 165}

\definecolor{greenOV}{RGB}{232, 245, 233}
\definecolor{greenI}{RGB}{200, 230, 201}
\definecolor{greenII}{RGB}{165, 214, 167}
\definecolor{greenIII}{RGB}{129, 199, 132}
\definecolor{greenIV}{RGB}{102, 187, 106}
\definecolor{greenV}{RGB}{76, 175, 80}
\definecolor{greenVI}{RGB}{67, 160, 71}
\definecolor{greenVII}{RGB}{56, 142, 60}
\definecolor{greenVIII}{RGB}{46, 125, 50}
\definecolor{greenIX}{RGB}{27, 94, 32}
\definecolor{greenAI}{RGB}{185, 246, 202}
\definecolor{greenAII}{RGB}{105, 240, 174}
\definecolor{greenAIV}{RGB}{0, 230, 118}
\definecolor{greenAVII}{RGB}{0, 200, 83}

\definecolor{lightgreenOV}{RGB}{241, 248, 233}
\definecolor{lightgreenI}{RGB}{220, 237, 200}
\definecolor{lightgreenII}{RGB}{197, 225, 165}
\definecolor{lightgreenIII}{RGB}{174, 213, 129}
\definecolor{lightgreenIV}{RGB}{156, 204, 101}
\definecolor{lightgreenV}{RGB}{139, 195, 74}
\definecolor{lightgreenVI}{RGB}{124, 179, 66}
\definecolor{lightgreenVII}{RGB}{104, 159, 56}
\definecolor{lightgreenVIII}{RGB}{85, 139, 47}
\definecolor{lightgreenIX}{RGB}{51, 105, 30}
\definecolor{lightgreenAI}{RGB}{204, 255, 144}
\definecolor{lightgreenAII}{RGB}{178, 255, 89}
\definecolor{lightgreenAIV}{RGB}{118, 255, 3}
\definecolor{lightgreenAVII}{RGB}{100, 221, 23}

\definecolor{limeOV}{RGB}{249, 251, 231}
\definecolor{limeI}{RGB}{240, 244, 195}
\definecolor{limeII}{RGB}{230, 238, 156}
\definecolor{limeIII}{RGB}{220, 231, 117}
\definecolor{limeIV}{RGB}{212, 225, 87}
\definecolor{limeV}{RGB}{205, 220, 57}
\definecolor{limeVI}{RGB}{192, 202, 51}
\definecolor{limeVII}{RGB}{175, 180, 43}
\definecolor{limeVIII}{RGB}{158, 157, 36}
\definecolor{limeIX}{RGB}{130, 119, 23}
\definecolor{limeAI}{RGB}{244, 255, 129}
\definecolor{limeAII}{RGB}{238, 255, 65}
\definecolor{limeAIV}{RGB}{198, 255, 0}
\definecolor{limeAVII}{RGB}{174, 234, 0}

\definecolor{yellowOV}{RGB}{255, 253, 231}
\definecolor{yellowI}{RGB}{255, 249, 196}
\definecolor{yellowII}{RGB}{255, 245, 157}
\definecolor{yellowIII}{RGB}{255, 241, 118}
\definecolor{yellowIV}{RGB}{255, 238, 88}
\definecolor{yellowV}{RGB}{255, 235, 59}
\definecolor{yellowVI}{RGB}{253, 216, 53}
\definecolor{yellowVII}{RGB}{251, 192, 45}
\definecolor{yellowVIII}{RGB}{249, 168, 37}
\definecolor{yellowIX}{RGB}{245, 127, 23}
\definecolor{yellowAI}{RGB}{255, 255, 141}
\definecolor{yellowAII}{RGB}{255, 255, 0}
\definecolor{yellowAIV}{RGB}{255, 234, 0}
\definecolor{yellowAVII}{RGB}{255, 214, 0}

\definecolor{amberOV}{RGB}{255, 248, 225}
\definecolor{amberI}{RGB}{255, 236, 179}
\definecolor{amberII}{RGB}{255, 224, 130}
\definecolor{amberIII}{RGB}{255, 213, 79}
\definecolor{amberIV}{RGB}{255, 202, 40}
\definecolor{amberV}{RGB}{255, 193, 7}
\definecolor{amberVI}{RGB}{255, 179, 0}
\definecolor{amberVII}{RGB}{255, 160, 0}
\definecolor{amberVIII}{RGB}{255, 143, 0}
\definecolor{amberIX}{RGB}{255, 111, 0}
\definecolor{amberAI}{RGB}{255, 229, 127}
\definecolor{amberAII}{RGB}{255, 215, 64}
\definecolor{amberAIV}{RGB}{255, 196, 0}
\definecolor{amberAVII}{RGB}{255, 171, 0}

\definecolor{orangeOV}{RGB}{255, 243, 224}
\definecolor{orangeI}{RGB}{255, 224, 178}
\definecolor{orangeII}{RGB}{255, 204, 128}
\definecolor{orangeIII}{RGB}{255, 183, 77}
\definecolor{orangeIV}{RGB}{255, 167, 38}
\definecolor{orangeV}{RGB}{255, 152, 0}
\definecolor{orangeVI}{RGB}{251, 140, 0}
\definecolor{orangeVII}{RGB}{245, 124, 0}
\definecolor{orangeVIII}{RGB}{239, 108, 0}
\definecolor{orangeIX}{RGB}{230, 81, 0}
\definecolor{orangeAI}{RGB}{255, 209, 128}
\definecolor{orangeAII}{RGB}{255, 171, 64}
\definecolor{orangeAIV}{RGB}{255, 145, 0}
\definecolor{orangeAVII}{RGB}{255, 109, 0}

\definecolor{deeporangeOV}{RGB}{251, 233, 231}
\definecolor{deeporangeI}{RGB}{255, 204, 188}
\definecolor{deeporangeII}{RGB}{255, 171, 145}
\definecolor{deeporangeIII}{RGB}{255, 138, 101}
\definecolor{deeporangeIV}{RGB}{255, 112, 67}
\definecolor{deeporangeV}{RGB}{255, 87, 34}
\definecolor{deeporangeVI}{RGB}{244, 81, 30}
\definecolor{deeporangeVII}{RGB}{230, 74, 25}
\definecolor{deeporangeVIII}{RGB}{216, 67, 21}
\definecolor{deeporangeIX}{RGB}{191, 54, 12}
\definecolor{deeporangeAI}{RGB}{255, 158, 128}
\definecolor{deeporangeAII}{RGB}{255, 110, 64}
\definecolor{deeporangeAIV}{RGB}{255, 61, 0}
\definecolor{deeporangeAVII}{RGB}{221, 44, 0}

\definecolor{brownOV}{RGB}{239, 235, 233}
\definecolor{brownI}{RGB}{215, 204, 200}
\definecolor{brownII}{RGB}{188, 170, 164}
\definecolor{brownIII}{RGB}{161, 136, 127}
\definecolor{brownIV}{RGB}{141, 110, 99}
\definecolor{brownV}{RGB}{121, 85, 72}
\definecolor{brownVI}{RGB}{109, 76, 65}
\definecolor{brownVII}{RGB}{93, 64, 55}
\definecolor{brownVIII}{RGB}{78, 52, 46}
\definecolor{brownIX}{RGB}{62, 39, 35}

\definecolor{grayOV}{RGB}{250, 250, 250}
\definecolor{grayI}{RGB}{245, 245, 245}
\definecolor{grayII}{RGB}{238, 238, 238}
\definecolor{grayIII}{RGB}{224, 224, 224}
\definecolor{grayIV}{RGB}{189, 189, 189}
\definecolor{grayV}{RGB}{158, 158, 158}
\definecolor{grayVI}{RGB}{117, 117, 117}
\definecolor{grayVII}{RGB}{97, 97, 97}
\definecolor{grayVIII}{RGB}{66, 66, 66}
\definecolor{grayIX}{RGB}{33, 33, 33}

\definecolor{bluegrayOV}{RGB}{236, 239, 241}
\definecolor{bluegrayI}{RGB}{207, 216, 220}
\definecolor{bluegrayII}{RGB}{176, 190, 197}
\definecolor{bluegrayIII}{RGB}{144, 164, 174}
\definecolor{bluegrayIV}{RGB}{120, 144, 156}
\definecolor{bluegrayV}{RGB}{96, 125, 139}
\definecolor{bluegrayVI}{RGB}{84, 110, 122}
\definecolor{bluegrayVII}{RGB}{69, 90, 100}
\definecolor{bluegrayVIII}{RGB}{55, 71, 79}
\definecolor{bluegrayIX}{RGB}{38, 50, 56}
\definecolor{bluegrayX}{RGB}{17, 23, 26}

\definecolor{myACMBlue}{cmyk}{1,0.1,0,0.1}
\definecolor{myACMYellow}{cmyk}{0,0.16,1,0}
\definecolor{myACMOrange}{cmyk}{0,0.42,1,0.01}
\definecolor{myACMRed}{cmyk}{0,0.90,0.86,0}
\definecolor{myACMLightBlue}{cmyk}{0.49,0.01,0,0}
\definecolor{myACMGreen}{cmyk}{0.20,0,1,0.19}
\definecolor{myACMPurple}{cmyk}{0.55,1,0,0.15}
\definecolor{myACMDarkBlue}{cmyk}{1,0.58,0,0.21}

\newcommand{\link}[1]{{\href{#1}{\color{blueVI}\textbf{\texttt{#1}}}}}
\newcommand{\myhref}[2]{{\href{#1}{\color{blueVI}\textbf{#2}}}}

\newcommand{\mypar}[1]{\vspace{3pt}\textbf{{#1}}}

\newcommand{\figpart}[1]{\textcolor{myACMPurple}{#1}}

\newcommand{\headertopspace}{\vspace{-1pt}}
\newcommand{\headerbottomspace}{\vspace{-1pt}}

\newcommand{\tool}{\textsc{Wordflow}}

\newcommand{\editorview}{\textit{Editor View}}
\newcommand{\localview}{\textit{Personal Prompt Library}}
\newcommand{\communityview}{\textit{Community Prompt Hub}}
\newcommand{\settingview}{\textit{Setting Panel}}
\newcommand{\toolbar}{\textit{Floating Toolbar}}
\newcommand{\promptcard}{\textit{Prompt Card}}
\newcommand{\prompteidtor}{\textit{Prompt Editor}}
\newcommand{\promptviewer}{\textit{Prompt Viewer}}

\definecolor{soulorange}{RGB}{255, 212, 153}
\definecolor{soulgray}{RGB}{220, 220, 220}
\definecolor{soulgraylight}{RGB}{235, 235, 235}
\definecolor{soulred}{RGB}{252, 217, 218}
\definecolor{soulbluelight}{RGB}{208, 233, 253}
\definecolor{souldorangelight}{RGB}{254, 234, 212}
\colorlet{soulblue}{blueV!30}

\newcommand{\inlinefig}[2]{\raisebox{-2pt}{\includegraphics[height=#1pt]{figures/#2}}}

\definecolor{tagbordercolor}{rgb}{0.8, 0.8, 0.8}
\definecolor{tagbgcolor}{rgb}{0.9, 0.9, 0.9}
\definecolor{lightgray}{RGB}{247, 247, 247}
\definecolor{midgray}{RGB}{179, 179, 179}

\newcommand*\textprompt[1]{\textcolor{tealVII}{``#1''}}
\newcommand*\textoutput[1]{\textcolor{lightblueVIII}{``#1''}}

\newcommand*\textprompttt[1]{\textcolor{tealVII}{\texttt{#1}}}
\newcommand*\textoutputtt[1]{\textcolor{lightblueVIII}{\texttt{#1}}}

\definecolor{tagbgcolor}{rgb}{1, 1, 1}

\definecolor{boxyellow}{RGB}{206, 171, 1}
\definecolor{boxgreen}{RGB}{14, 152, 136}
\definecolor{boxblue}{RGB}{77, 167, 223}

\setul{0.45ex}{0.2ex}

\begin{document}

\title{\tool{}: Social Prompt Engineering for Large Language Models}

\settopmatter{authorsperrow=4}

\author{Zijie J. Wang}
\orcid{0000-0003-4360-1423}
\affiliation{%
  \institution{Georgia Tech}
  \city{Atlanta}
  \state{Georgia}
  \country{USA}
}

\author{Aishwarya Chakravarthy}
\orcid{0009-0003-4481-2844}
\affiliation{%
  \institution{Georgia Tech}
  \city{Atlanta}
  \state{Georgia}
  \country{USA}
}

\author{David Munechika}
\orcid{0000-0002-3643-6899}
\affiliation{%
  \institution{Georgia Tech}
  \city{Atlanta}
  \state{Georgia}
  \country{USA}
}

\author{Duen Horng Chau}
\orcid{0000-0001-9824-3323}
\affiliation{%
  \institution{Georgia Tech}
  \city{Atlanta}
  \state{Georgia}
  \country{USA}
}
\renewcommand{\shortauthors}{Zijie J. Wang, et al.}

\begin{abstract}
  Large language models (LLMs) require well-crafted prompts for effective use.
  Prompt engineering, the process of designing prompts, is challenging, particularly for non-experts who are less familiar with AI technologies.
  While researchers have proposed techniques and tools to assist LLM users in prompt design, these works primarily target AI application developers rather than non-experts.
  To address this research gap, we propose social prompt engineering, a novel paradigm that leverages social computing techniques to facilitate collaborative prompt design.
  To investigate social prompt engineering, we introduce \tool{}, an open-source and social text editor that enables everyday users to easily create, run, share, and discover LLM prompts.
  Additionally, by leveraging modern web technologies, \tool{} allows users to run LLMs locally and privately in their browsers.
  Two usage scenarios highlight how our tool's incorporation of social prompt engineering can enhance laypeople's interactions with LLMs.
  \tool{} is publicly accessible at \link{https://poloclub.github.io/wordflow}.
\end{abstract}

\ccsdesc[500]{Computing methodologies~Machine learning}
\ccsdesc[500]{Human-centered computing~Human computer interaction (HCI)}
\ccsdesc[500]{Human-centered computing~Interactive systems and tools}

\keywords{Prompt Engineering, Large Language Model, Machine Learning}

\begin{teaserfigure}
  \centering
  \includegraphics[width=0.82\textwidth]{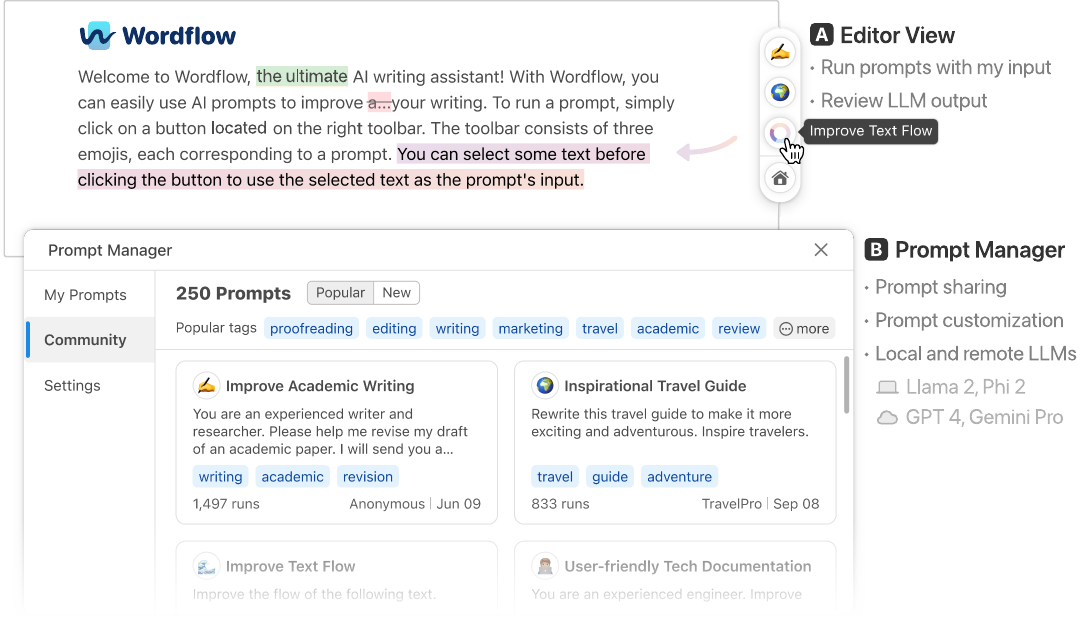}
  \vspace{-8pt}
  \caption{
    \tool{} is an open-source social prompt engineering tool to help everyday users create, run, share, and discover prompts for large language models (LLMs).
    \textbf{(A) The \editorview{}} offers an easy-to-use text editing interface, allowing users to run an LLM prompt using the selected text as input by simply clicking on a button and examine the changes made by LLMs.
    \textbf{(B) The \textit{Prompt Manager}} enables users to edit and curate prompts, adjust LLM settings, and share their prompts with the community.
  }
  \vspace{8pt}
  \Description{Screenshot of \tool{}.}
  \label{fig:teaser}
\end{teaserfigure}

\maketitle
\pagestyle{plain}

\headertopspace{}
\section{Introduction}
\headerbottomspace{}

Recently, there has been a surge in the popularity of large language models (LLMs) such as GPT-4~\cite{openaiGPT4TechnicalReport2023}, Gemini~\cite{teamGeminiFamilyHighly2023}, and Llama 2~\cite{touvronLlamaOpenFoundation2023}.
These pre-trained artificial intelligence (AI) models demonstrate a diverse array of capabilities that are continually being discovered, including summarization, question-answering, creative writing, and translation~\cite{bommasaniOpportunitiesRisksFoundation2022,srivastavaImitationGameQuantifying2022a}.
To instruct these general-purpose LLMs to perform specific tasks, users need to provide them with \textit{prompts}---text instructions and examples of desired outputs~\cite{brownLanguageModelsAre2020, lewisRetrievalAugmentedGenerationKnowledgeIntensive2021}.
These prompts serve as background contexts and guides for LLMs to generate text that aligns with users' objectives.
Prompting enables users to employ LLMs for various tasks with plain language; in fact, well-crafted prompts can make general-purpose LLMs outperform specialized AI models~\cite{noriCanGeneralistFoundation2023}.

Designing effective prompts, known as \textit{prompt engineering}, poses significant challenges for LLM users~\cite{liuPretrainPromptPredict2023,jiangPromptMakerPromptbasedPrototyping2022}.
LLM users often rely on trial and error and employ unintuitive patterns, such as adding \textprompt{think step by step}~\cite{kojimaLargeLanguageModels2022} to their prompts, to successfully instruct LLMs.
Prompt engineering, despite its name, is considered an art~\cite{parameswaranRevisitingPromptEngineering2023} and is even compared to wizards learning ``magic spells''~\cite{harwellTechHottestNew2023, willisonStableDiffusionBreaks2022}.
Prompt writers may not fully understand why certain prompts work, but they still add them to their ``spell books.''
Furthermore, prompting is especially challenging for \textit{non-AI-experts}, who are often confused about getting started and lack sufficient guidance and training on LLMs and prompting~\cite{zamfirescu-pereiraWhyJohnnyCan2023, zhouInstructPipeBuildingVisual2023}.

There is a growing body of research on helping users prompt LLMs.
Researchers propose instruction tuning~\cite{chungScalingInstructionFinetunedLanguage2022, weiFinetunedLanguageModels2022} and reinforcement learning from human feedback~\cite{ouyangTrainingLanguageModels2022, stiennonLearningSummarizeHuman2020} to align a model's output with users' intent.
Prompt techniques~\cite[e.g.,][]{brownLanguageModelsAre2020, weiChainofthoughtPromptingElicits2022, noriCanGeneralistFoundation2023} are introduced to improve LLMs' performance on complex tasks.
Libraries~\cite{chaseLangChainBuildingApplications2022, lundbergGuidanceGuidanceLanguage2023} and interactive tools~\cite[e.g.,][]{wuAIChainsTransparent2022, arawjoChainForgeVisualToolkit2023, jiangPromptMakerPromptbasedPrototyping2022, strobeltInteractiveVisualPrompt2022, fiannacaProgrammingProgrammingLanguage2023} have also been developed to streamline the prompt crafting process.
However, these techniques and tools primarily cater to AI application developers who use LLMs to build AI applications~(e.g., chatbot applications), overlooking non-expert users who use LLMs for everyday tasks~(e.g., checking emails for grammar errors).
To bridge this critical research gap, we propose \textit{social prompt engineering}, a novel paradigm that leverages social computing techniques to facilitate collaborative prompt designs.
\textbf{We contribute:}

\begin{itemize}[topsep=5pt, itemsep=0mm, parsep=1mm, leftmargin=10pt]
      \item \textbf{\tool{}, the first social and customizable text editor} that empowers everyday users to create, run, share, and discover LLM prompts~(\autoref{fig:teaser}).
            It features a direct manipulation text editing interface for applying LLM prompts to transform existing text, such as proofreading and translation, or generate new text, such as creative writing.
            Users can easily customize prompts and LLM settings, share prompts with the community, and copy community prompts~(\autoref{sec:design}).
            Two usage scenarios highlight how \tool{} and social prompt engineering can enhance users' interactions with LLMs~(\autoref{sec:scenario}).
            Finally, we discuss future research directions for integrating workflows, fostering user engagement and responsible AI, and evaluation~(\autoref{sec:discussion}).

      \item \textbf{An open-source\footnote{\tool{} code: \link{https://github.com/poloclub/wordflow}}, web-based implementation} that lowers the barrier for everyday users in designing effective prompts and applying LLMs to their daily tasks.
            By leveraging modern web technologies, such as WebGPU~\cite{mdnWebGPUAPIWeb2023,teamMLCLLM2023}, our tool enables users to run cutting-edge LLMs locally without the need for dedicated backend servers or external LLM API services~(\autoref{sec:design:implementation}).
            Additionally, we offer an open-source implementation to help future designers and researchers adopt \tool{} for exploring and developing future user interfaces for LLMs.
            To see a demo of \tool{}, visit \link{https://youtu.be/3dOcVuofGVo}.
\end{itemize}

\noindent{}Using \tool{} as a design probe, we plan to release it to the public and collect usage data to assess the effectiveness of social prompt engineering and investigate how users collaboratively craft prompts.
We hope our work will inspire the design, research, and development of collaborative interfaces that help everyone more easily and effectively use LLMs.

\setlength{\belowcaptionskip}{-6pt}
\setlength{\abovecaptionskip}{2pt}
\begin{figure*}[tb]
      \includegraphics[width=1\linewidth]{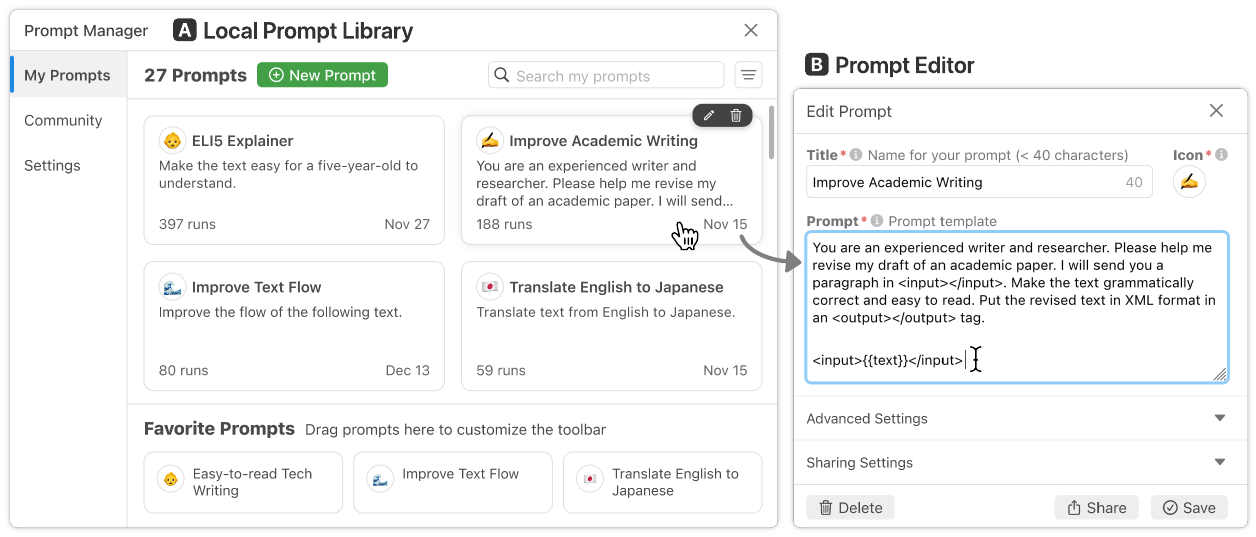}
      \caption[]{
            With \tool{}, users can easily manage and customize their prompts.
            \textbf{(A) The \localview{}} provides an overview of local prompts, allowing users to search, sort, and customize the quick-action prompt toolbar in the \editorview{}.
            \textbf{(B) The \prompteidtor{}}, activated by clicking a \promptcard{}, employs progressive disclosure to help users edit basic prompt information, advanced settings (e.g., output parsing rules and LLM temperature), and sharing configurations.
      }
      \Description{
            A screenshot of the \localview{}.
      }
      \label{fig:library}
\end{figure*}
\setlength{\belowcaptionskip}{0pt}
\setlength{\abovecaptionskip}{12pt} %
\section{Background \& Related Work}
\headerbottomspace{}

\mypar{Using LLMs by prompting.}
LLMs are trained to generate plausible output by continuing an input text, also known as a prompt.
These models are pre-trained on billions of text samples from the internet, enabling them to perform various tasks specified in the prompt through auto-completion.
For example, given a prompt with translation examples like \textprompt{English: Hello. Spanish: Hola. English: Good luck! Spanish:}, an LLM would auto-complete it with the translation \textoutput{¡Buena suerte!}, making LLMs useful for translation.
Recently, LLMs have been fine-tuned using user prompts to simplify prompting~\cite{chungScalingInstructionFinetunedLanguage2022,weiFinetunedLanguageModels2022,ouyangTrainingLanguageModels2022}.
To instruct an instruction-tuned LLM to translate English to Spanish, one can prompt it with \textprompt{Translate the following English sentence to Spanish: Good luck!}, and the LLM would output \textoutput{¡Buena suerte!}.

\mypar{Challenges of prompt engineering.}
The accuracy of LLMs depends heavily on the prompts~\cite{noriCanGeneralistFoundation2023,liuPretrainPromptPredict2023}.
However, prompt engineering, the process of crafting effective prompts, is difficult.
Researchers have shown slight wording changes in the prompt can significantly impact LLM accuracy~\cite{zamfirescu-pereiraWhyJohnnyCan2023}.
A prompt's effectiveness can vary greatly across different models~\cite{woolfProblemLangChain2023}.
The LLM community has discovered unintuitive prompting patterns that can greatly enhance LLMs' performance, such as priming the LLM with phrases like \textprompt{you are a translation expert} and improving LLM's reasoning capability with \textprompt{think step by step}~\cite{kojimaLargeLanguageModels2022} or chain-of-thought prompting~\cite{weiFinetunedLanguageModels2022}.
The brittleness of prompts and unintuitive prompting patterns make it difficult for LLM users, especially everyday users unfamiliar with AI, to write effective prompts~\cite{zamfirescu-pereiraWhyJohnnyCan2023}.

\mypar{Addressing prompt engineering challenges.}
Researchers have proposed libraries such as \textsc{LangChain}~\cite{chaseLangChainBuildingApplications2022}, \textsc{Guidance}~\cite{lundbergGuidanceGuidanceLanguage2023}, and \textsc{Outlines}~\cite{willardEfficientGuidedGeneration2023} to help users write prompts programmatically and control the structure of an LLM's output.
By formulating prompting as programming, researchers propose integrated development environment (IDE) features that help users edit~\cite{fiannacaProgrammingProgrammingLanguage2023} and unit test prompts~\cite{strobeltPromptTesterQuick2023}.
Similarly, \textsc{CoPrompt}~\cite{fengCoPromptSupportingPrompt2023} introduces a collaborative editor for multiple programmers to write prompts simultaneously.
AI prototyping tools like \textsc{PromptMaker}~\cite{jiangPromptMakerPromptbasedPrototyping2022}, \textsc{Google AI Studio}~\cite{googleGoogleAIStudio2023}, \textsc{OpenAI Playground}~\cite{openaiOpenAIPlayground2023}, and \textsc{PartyRock}~\cite{amazonPartyRockEveryoneCan2023} allow users to rapidly write and run prompts.
Leveraging visual programming techniques, \textsc{AI Chains}~\cite{wuAIChainsTransparent2022}, \textsc{PromptChainer}~\cite{wuPromptChainerChainingLarge2022}, \textsc{Prompt Sapper}~\cite{chengPromptSapperLLMEmpowered2023}, and \textsc{ChainForge}~\cite{arawjoChainForgeVisualToolkit2023} enable AI application developers to visually design and test complex prompts.
Similarly, \textsc{PromptIDE}~\cite{strobeltInteractiveVisualPrompt2022}, \textsc{PromptAID}~\cite{mishraPromptAidPromptExploration2023}, and \textsc{Prompterator}~\cite{sucikPrompteratorIterateEfficiently2023} employ mixed-initiative and interactive visualization techniques to help LLM users brainstorm and refine prompts.
These existing tools function as IDEs that help \textit{AI developers} craft prompts that will later be integrated into other applications.
In contrast, \tool{} aims to serve as a runtime interface for \textit{everyday users}, who act as both the prompt engineers and direct users of their prompts, and may not be well-versed in AI technologies.

\mypar{Social prompt engineering.}
Online communities, including Promptstacks~\cite{promptstacksPromptstacksYourPrompt2023}, ChatGPT Prompt Genius~\cite{redditChatGPTPromptGenius2023}, and ShareGPT~\cite{ecclestonShareGPTShareYour2022}, serve as platforms for prompt creators to share tips, collaborate, and stay updated on AI advancements.
User prompts from social media have been scraped to create prompt datasets for AI model development~\cite{wangDiffusionDBLargescalePrompt2023}.
Online prompt marketplaces, such as PromptBase~\cite{promptbasePromptBasePromptMarketplace2023}, PromptHero~\cite{promptheroPromptHeroSearchPrompts2023} and ChatX~\cite{chatxChatXChatGPTDALL2023}, have emerged to allow users to buy and sell prompts for generative models.
Midjourney's Discord server~\cite{holzMidjourneyExploringNew2022} allows users to run and share prompts for text-to-image generative models, with dedicated sections for prompt critique and improvement~\cite{oppenlaenderTaxonomyPromptModifiers2022}.
Building on the design of these communities, \tool{} provides an easy-to-use interface that unifies creating, running, sharing, and discovering LLM prompts.
The most relevant related work is \textsc{PromptSource}~\cite{bachPromptSourceIntegratedDevelopment2022}, an IDE for AI researchers and developers to write and share LLM prompts.
\textsc{PromptSource} targets AI experts using LLMs for natural language processing tasks on datasets (such as data annotation), and it requires users to provide a dataset.
In comparison, \tool{} targets everyday users using LLMs for daily tasks, such as grammar checking, without the need to provide any dataset. %
\headertopspace{}
\section{System Design \& Implementation}
\label{sec:design}
\headerbottomspace{}

\tool{} is an interactive tool that empowers everyday users to easily create, run, share, and discover LLM prompts.
It provides an easy-to-use interface that unifies prompt creation, execution, and sharing.
It tightly integrates four views: the \editorview{}~(\autoref{sec:design:editor}), where users can write text, run LLM prompts, and inspect changes made by LLMs;
the \localview{}~(\autoref{sec:design:local}), offering a prompt manager for creating, editing, and curating prompts locally;
the \communityview{}~(\autoref{sec:design:local}), enabling users to explore and search for the latest and popular prompts shared by the community;
and the \settingview{}, where users can configure LLMs to run their prompts with remote or local models~(\autoref{sec:design:implementation}).

\headertopspace{}
\subsection{Editor View}
\label{sec:design:editor}
\headerbottomspace{}

\setlength{\columnsep}{5pt}%
\setlength{\intextsep}{0pt}%
\begin{wrapfigure}{r}{30pt}
  \centering
  \includegraphics[width=30pt]{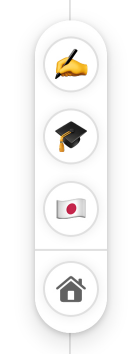}
  \vspace{-12pt}
  \label{fig:llm-menu}
  \Description{Screenshot of Wordflow.}
\end{wrapfigure}
When users open \tool{} in their browser or its mobile and desktop progressive web app, they are presented with the \editorview{}~(\autoref{fig:teaser}\figpart{A}).
This view shows a familiar text editor interface with a \toolbar{} anchored on the right.
Users can type or paste text into the editor.
The \toolbar{} consists of three prompt buttons and a home button~(shown on the right).
Each prompt button is represented by an emoji icon and corresponds to a prompt template.
Users can click the prompt button to run its prompt using the current paragraph as the input text.
If a user has selected some text, the selected text is used as the input for the prompt.
Users can also click the home button~\inlinefig{10}{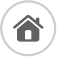} to open a pop-up window that contains the \localview{}~(\autoref{sec:design:local}, \autoref{fig:library}\figpart{A}), the \communityview{}~(\autoref{sec:design:community}, \autoref{fig:teaser}\figpart{B}), and the \settingview{}~(\autoref{fig:llm-selection}).

\mypar{Prompt input templating.}
In \tool{}, a prompt template includes pre-defined prefix text and a placeholder for the input text.
For example, the prefix text can be \textprompt{Improve the flow of the following text}.
The input placeholder in the template serves as a variable that will be substituted with the selected text from the editor.
Inspired by popular prompting tools such as \textsc{LangChain}~\cite{chaseLangChainBuildingApplications2022} and \textsc{PromptMaker}~\cite{jiangPromptMakerPromptbasedPrototyping2022}, our tool supports basic prompt templating.
Users can include a special string \textprompttt{\{\{text\}\}} in their prompt template to represent the input placeholder~(\autoref{fig:library}\figpart{B}), which will be replaced with the selected text from the editor before running the prompt.
If the user does not include the string \textprompttt{\{\{text\}\}} in the template, the input text will be appended to the prompt template.

\setlength{\belowcaptionskip}{-5pt}
\setlength{\abovecaptionskip}{2pt}
\begin{figure}[tb]
  \centering
  \includegraphics[width=0.8\linewidth]{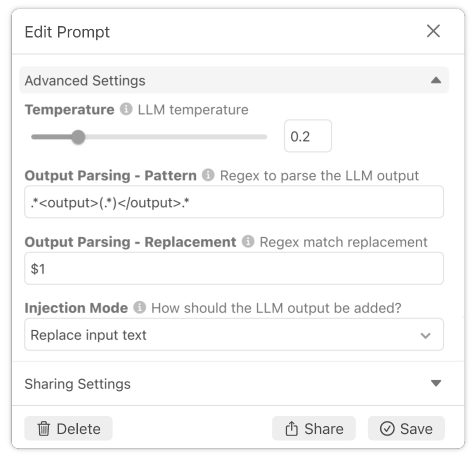}
  \caption{
    The \prompteidtor{} allows users to easily configure LLM settings such as temperature and output parsing rules.
  }
  \label{fig:advanced-settings}
  \Description{Screenshot of Wordflow.}
\end{figure}
\setlength{\belowcaptionskip}{0pt}
\setlength{\abovecaptionskip}{12pt}

\mypar{Prompt output parsing.}
To run users' prompts, \tool{} supports remote LLM API services, such as GPT 4~\cite{openaiGPT4TechnicalReport2023} and Gemini~\cite{teamGeminiFamilyHighly2023} API services provided by OpenAI and Google, as well as local open-source models, such as Llama 2~\cite{touvronLlamaOpenFoundation2023} and Phi 2~\cite{abdinPhi2SurprisingPower2023}.
Users can set their preferred models in the \settingview{}~(\autoref{fig:llm-selection}).
After receiving the output from the LLM API service or local model, the \editorview{} applies Myer's diffing algorithm~\cite{myersNDDifferenceAlgorithm1986, fraserDiffmatchpatchHghperformanceLibrary2012} to compare the output text with the input text.
It then highlights the changes made by the LLM (e.g., addition, replacement, and deletion) using different text background colors~(\autoref{fig:teaser}\figpart{A}).
Users can click on the highlighted text to accept or reject the changes.
Inspired by \textsc{LangChain}, \tool{} allows users to add \textit{optional} output parsing rules to a prompt by writing regular expression (regex) text~(\autoref{fig:advanced-settings}).
For example, a user can prompt LLMs to structure the output in XML format (recommended by prompt engineering guidelines~\cite{anthropicIntroductionPromptDesign2023}), such as \textprompt{Improve the flow of the following text. Put the rewritten text in an XML tag \texttt{<output></output>}}.
The user can then add a regex pattern \textoutputtt{.*<output>(.*)</output>.*} and a replacement rule \textoutputtt{\$1} to parse the LLM's output before it is displayed in the \editorview{}.
This feature is useful for disregarding unrelated text in the LLM's output.
For instance, the output \textoutput{Sure, I can help you! \texttt{<output>}Over recent years...\texttt{</output>}} will be parsed as \textoutput{Over recent years...}.
Furthermore, users can configure the insertion mode for each prompt.
In \texttt{replace} mode, the input text is replaced with the LLM output, while in \texttt{append} mode, the LLM output is appended to the input text.

\headertopspace{}
\subsection{Personal Prompt Library}
\label{sec:design:local}
\headerbottomspace{}

After clicking the home button~\inlinefig{10}{icon-button-home}, users can open the \localview{} to manage their local prompts~(\autoref{fig:library}\figpart{A}).
This view organizes each prompt as a \promptcard{}, allowing users to easily search and sort prompts based on name, recency, and run count.
To change the prompts in the \toolbar{}~(\autoref{sec:design:editor}), users can simply drag a \promptcard{} into one of the three prompt slots located in the bottom row, each corresponding to a prompt button in the \toolbar{}.
To add or edit a prompt, users can click on the \inlinefig{10}{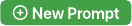} button or a \promptcard{} to open the \prompteidtor{}~(\autoref{fig:library}\figpart{B}).
The \prompteidtor{} comprises three forms: basic prompt information~(\autoref{fig:library}\figpart{B}), optional advanced settings~(\autoref{fig:advanced-settings}), and optional sharing settings.
In the basic prompt information section, users can configure the title, icon, and prompt template.
The advanced settings allow more experienced users to set the LLM temperature, output parsing rules, and insertion rules~(\autoref{fig:advanced-settings}).
To share a prompt with the community, users can provide a description, tags, and recommended LLM models in the sharing settings, and then click on the \inlinefig{10}{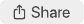} button.

\headertopspace{}
\subsection{Community Prompt Hub}
\label{sec:design:community}
\headerbottomspace{}

The \communityview{} enables users to easily browse and search for prompts shared by \tool{} users~(\autoref{fig:teaser}\figpart{B}).
Each community prompt is represented as a \promptcard{} and is associated with at least one tag.
Users can filter prompts by clicking on a tag and can also sort prompts based on recency and popularity (i.e., the number of times they have been run).
Additionally, users can view the most popular tags on the top row of this panel.
By clicking on a \promptcard{}, users can access the \promptviewer{}~(\autoref{fig:prompt-viewer}) to examine detailed information provided by the prompt creator, including the title, description, prompt template, and recommended LLM models.
Finally, users can click on the \inlinefig{9}{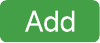} button to include a copy of the community prompt in their \localview{}~(\autoref{sec:design:local}), where they can run the prompt, make further refinements, and potentially share it again with the community.

\headertopspace{}
\subsection{Open-source Implementation}
\label{sec:design:implementation}
\headerbottomspace{}

\setlength{\columnsep}{10pt}%
\setlength{\intextsep}{-3pt}%
\begin{wrapfigure}{r}{120pt}
  \centering
  \includegraphics[width=90pt]{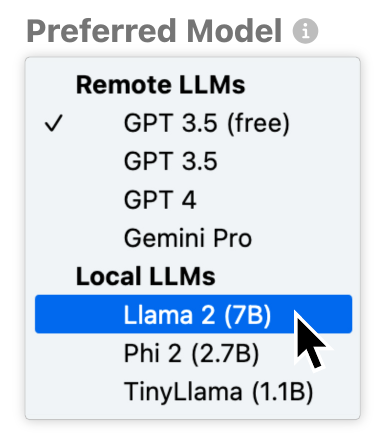}
  \vspace{-12pt}
  \caption{
    \tool{} supports remote and local LLMs.
  }
  \label{fig:llm-selection}
  \Description{Screenshot of Wordflow.}
\end{wrapfigure}

We implement \tool{} as a progressive web app using Web Components~\cite{mdnWebComponentsWeb2021} and LIT Element~\cite{googleLitSimpleFast2020} as the frontend framework.
Users can use \tool{} as a mobile or desktop app by saving it as a Safari Web App~\cite{appleUseSafariWeb2023} or a Chrome app~\cite{googleAddOpenChrome2022}.
\tool{} allows users to run LLMs through remote API services, such as GPT 4 provided by OpenAI, or directly run open-source LLMs, such as Llama 2, Phi 2, and TinyLlama~\cite{zhangTinyLlamaOpenSourceSmall2024}, in their browser~(\autoref{fig:llm-selection}).
We use Web LLM~\cite{teamMLCLLM2023} and WebGPU~\cite{mdnWebGPUAPIWeb2023} to implement on-device LLM inference.
In \tool{}, all local prompts are stored in the local persistent storage of the user's browser.
To enable users to share community prompts, we use Amazon API Gateway~\cite{amazonAmazonAPIGateway2023} and DynamoDB~\cite{amazonAmazonDynamoDBFast2023} as a backend.
Additionally, we provide a Google Doc add-on~(\autoref{fig:doc}) that allows Google Doc users to directly use \tool{} within their editor.
We \myhref{https://github.com/poloclub/wordflow}{open source} \tool{} as a collection of reusable interactive components that can be easily adopted by future researchers and designers in their interactive LLM projects. %
\headertopspace{}
\section{Usage Scenarios}
\label{sec:scenario}

\subsection{Improving Technical Writing}
\headerbottomspace{}

\setlength{\belowcaptionskip}{0pt}
\setlength{\abovecaptionskip}{5pt}
\begin{figure}[tb]
  \centering
  \includegraphics[width=0.9\linewidth]{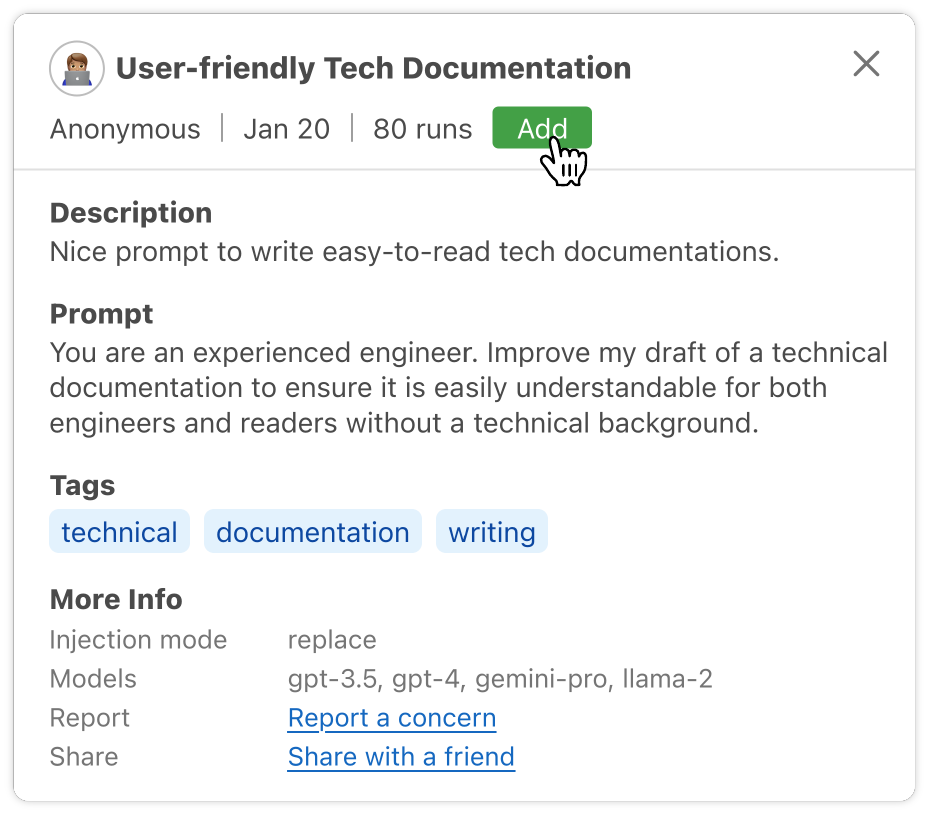}
  \caption{
    The \promptviewer{} shows detailed information about a community prompt.
    Users can click a button to copy this prompt into their \localview{}.
  }
  \label{fig:prompt-viewer}
  \Description{Screenshot of Wordflow.}
\end{figure}
\setlength{\belowcaptionskip}{0pt}
\setlength{\abovecaptionskip}{12pt}

As a recently graduated junior software developer, Wade has been struggling with writing API documentation and system architecture descriptions.
Specifically, Wade is unfamiliar with explaining technical concepts in simple language that can be easily understood by different colleagues such as developers, UX designers, and program managers.
One day, Wade came across a forum thread where developers were sharing LLM prompts that had helped them improve their technical documentation writing.
Wade had never thought about using LLM to assist him in his writing before.
Intrigued, he clicked on a \tool{} \myhref{https://poloclub.github.io/wordflow?prompt=b131cdb7-558e-5845-9b83-f877f5718b66}{prompt link} shared in a popular comment on the thread.
The link opened \tool{} in a new tab, displaying the \communityview{} along with a pop-up showing a community prompt~(\autoref{fig:prompt-viewer}).
Wade found the prompt and its description to be suitable for his writing tasks, so he clicked on the \inlinefig{9}{icon-button-add} button to copy this community prompt to his local library.

Wade decided to try out this prompt to improve his writing.
He opened the \localview{} and dragged the newly added prompt into one of the Favorite Prompts slots~(\autoref{fig:library}\figpart{A}), and the prompt appeared in the \toolbar{} in the \editorview{}~(\autoref{fig:teaser}\figpart{A}).
Wade copied a paragraph from the API documentation that he was working on.
However, before clicking on the prompt button in the \toolbar{}, Wade suddenly remembered that his company prohibits employees from using LLM services (e.g., ChatGPT and Bard) with work materials, as a measure to safeguard trade secrets and sensitive information.
Upon reviewing the documentation of \tool{}, Wade discovered that \tool{} supports running local LLMs directly in browsers without sending any data to third-party services (e.g., OpenAI and Google).
Therefore, he configured the LLM model to Llama 2, a local LLM model, in the \settingview{} before running the prompt on his writing.
Then, he observed the changes made by the LLM model, which were highlighted in the \editorview{}, and found the new paragraph to be much easier to read.
After using this prompt for a few days, Wade shared the prompt link on his company's mailing list, and more developers from his company began to use it to improve technical writing.

\headertopspace{}
\subsection{Customizing Translation Styles}
\headerbottomspace{}

\setlength{\belowcaptionskip}{-10pt}
\setlength{\abovecaptionskip}{5pt}
\begin{figure}[tb]
  \centering
  \includegraphics[width=\linewidth]{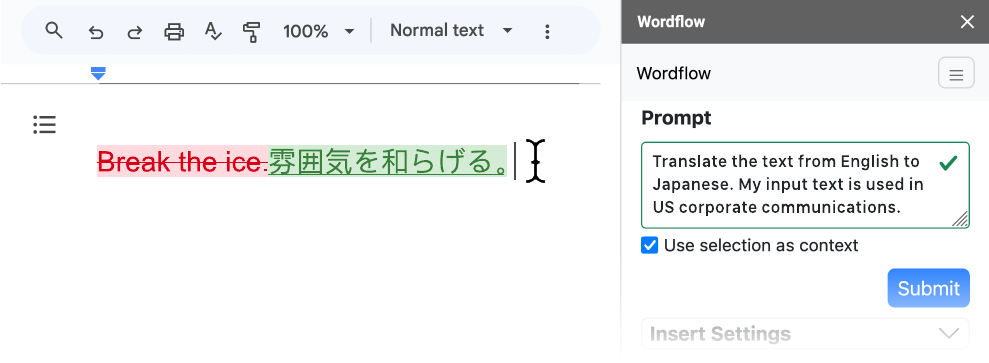}
  \caption{
    Google Doc users can directly use \tool{}'s add-on to apply prompts to text within their Google Doc documents.
  }
  \label{fig:doc}
  \Description{Screenshot of Wordflow.}
\end{figure}
\setlength{\belowcaptionskip}{0pt}
\setlength{\abovecaptionskip}{12pt}

\begin{CJK*}{UTF8}{min}
  Ember, a senior manager in a US financial firm, has faced a new challenge since her company started collaborating with a Japanese counterpart.
  Due to the absence of their Japanese translator, Ember has resorted to using translation software to communicate with the managers from the Japanese company.
  However, she has noticed that the translations generated by the software occasionally lead to confusion among her Japanese colleagues.
  For example, the software translated the English idiom ``break the ice'' to ``氷を砕く,'' which means ``destroy the ice'' instead of her intended meaning of ``relieving tension when people interact for the first time.''

  Due to the recent popularity of LLMs, Ember decided to try using them to translate her documents from English to Japanese.
  As she writes in Google Docs, she explored the Google Doc Marketplace for an AI add-on and came across \tool{}.
  Upon installation, she opened the \communityview{}~(\autoref{fig:teaser}\figpart{B}) and selected the tag \inlinefig{9}{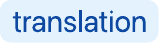}, which showed various popular translation prompts.
  She found a \myhref{https://poloclub.github.io/wordflow?prompt=49f7307e-90a3-5937-a63f-6dd42057c4d6}{prompt} titled ``Translate English to Japanese.''

  After adding this prompt to her library, she tried to run it with the input ``break the ice''.
  However, \tool{} appended the incorrect translation ``氷を砕く'' to her document.
  Drawing from her previous experience interacting with ChatGPT, Ember decided to edit the prompt and provide additional instructions to guide the LLM model in considering her translation context.
  She opened the \editorview{}~(\autoref{fig:library}\figpart{B}). and added a new sentence to the translation prompt: \textprompt{My input text is used in US corporate communications}~(\autoref{fig:doc}~\figpart{Right}).
  Running the prompt again, \tool{} generated a more suitable translation ``雰囲気を和らげる,'' which means ``ease the atmosphere''~(\autoref{fig:doc}~\figpart{Left}).
  Ember back-translated the translation to English using her other translation software and felt more confident in continuing to use this prompt for future translations.
  Finally, to help other people who need to translate English to Japanese in business settings, she shared her \myhref{https://poloclub.github.io/wordflow?prompt=7dff88cf-ff9a-587f-a3d1-f436301f72a2}{updated prompt} with the community by clicking on the \inlinefig{10}{icon-button-share} button~(\autoref{fig:library}\figpart{B}).

\end{CJK*} %
\headertopspace{}
\section{Discussion \& Future Work}
\label{sec:discussion}
\headerbottomspace{}

While \tool{} can help users create, run, share, and discover prompts, the current system can be improved in terms of workflow integration and social system design.
Finally, we plan to conduct a usage log study to evaluate social prompt engineering.

\mypar{Fitting into user workflows.}
The current version of \tool{} requires users to copy and paste their input text into a webpage or Google Doc.
To minimize disruption to users' workflow (e.g., drafting an email, replying messages, or editing a PowerPoint), future researchers can make \tool{} \textit{in situ} and ubiquitous.
For example, Wordflow can be directly integrated into an operating system, running a prompt when users select text and trigger a keyboard shortcut.
\tool{} supports both external LLM API services and on-device LLMs for running users' prompts~(\autoref{sec:design:implementation}).
With recent advancements in machine learning compilation~\cite{chenTVMAutomatedEndtoEnd2018} and model compression~\cite{hohmanModelCompressionPractice2023}, we see great potential for on-device LLMs.
Local LLMs allow users to avoid sending sensitive data to external services, reduce API costs, and use LLMs without network access.
To enhance the usability and development experience of on-device LLM tools, researchers can explore integrating local LLMs into the operating system.
This integration would enable various AI tools to run LLMs as ``system functions,'' eliminating the need for redundant LLM installations within each tool.

\mypar{Promoting user engagement.}
\tool{} is the first social prompt engineering tool to help everyday users create, run, and share prompts.
There are great research opportunities to enhance user interaction with LLMs by leveraging social computing techniques.
For example, future researchers can draw inspiration from gaming social platforms, such as Steam Community~\cite{sharmaExploringGamersCrowdsourcing2021} and Pokémon GO forums~\cite{loriaComparingStructuresCharacteristics2021}, where gamers engage in research and share strategies to overcome in-game challenges.
By comparing prompting LLMs to fighting game bosses, we can explore the design of social systems that motivate users to research and exchange prompting techniques.
To incentivize user participation in prompt sharing, researchers can explore \textit{intrinsic motivations}, such as designing an enjoyable social system~\cite{bakiciComparisonCrowdsourcingPlatforms2020}, and \textit{extrinsic motivations}, such as virtual rewards and reputation systems~\cite{wuMotivationSustainedParticipation2020, gohPerceptionsVirtualReward2017}.
Lastly, \tool{} users can filter and sort prompts by tags, recency, and popularity.
Researchers can explore using social media ranking techniques to recommend relevant community prompts to users~\cite{sarwarItembasedCollaborativeFiltering2001}.

\mypar{Fostering responsible AI practices.}
It has never been more critical to address the potential harms associated with LLMs~\cite{weidingerSociotechnicalSafetyEvaluation2023}.
Social prompt engineering presents interesting opportunities and challenges for responsible AI.
On the one hand, social systems like \tool{} can enable users to share prompting techniques~\cite{askellGeneralLanguageAssistant2021} to mitigate potential harms.
On the other hand, without content moderation, users can easily disseminate harmful prompts, such as a misinformation generator~\cite{hanleyMachineMadeMediaMonitoring2023}.
In \tool{}, users can report harmful prompts, and we will diligently monitor and moderate community prompts.
Future researchers can explore social system designs that promote responsible prompting and develop methods to detect potentially harmful prompts.

\mypar{Planned usage log study.}
Using \tool{} as a research instrument, we plan to conduct a usage log study to evaluate social prompt engineering and investigate two research questions:
\begin{enumerate}[topsep=2pt, itemsep=0mm, parsep=1mm, leftmargin=32pt, label={RQ\arabic*.}, ref=RQ\arabic*]
  \item \label{item:q1} How does social prompt engineering help everyday users craft prompts?
  \item \label{item:q2} What do everyday users use LLMs for, and how do they write prompts?
\end{enumerate}
To answer RQ1, we will examine the evolution of prompts by analyzing the usage patterns of copying community prompts and the modifications made to them before re-sharing these prompts.
To answer RQ2, we will conduct a mixed-method analysis of community prompts to synthesize popular use cases of LLMs and prompting patterns.
Our institution's IRB has approved our user study, and we will start data collection after refining the design of \tool{}.
\headertopspace{}
\section{Conclusion}
\label{sec:conclusion}
\headerbottomspace{}

As LLMs are increasingly being used by everyday users in their daily tasks, it is critical to help them write and run prompts easily.
In this work, we present social prompt engineering, a new paradigm that leverages social computing techniques to facilitate collaborative prompt design.
To investigate social prompt engineering, we design and develop \tool{}, an open-source and social text editor empowering users to easily create, run, share, and discover LLM prompts.
Two usage scenarios highlight social prompt engineering and \tool{} can assist everyday users in interacting with LLMs.
We discuss our ongoing work and future research directions.
We hope our work will inspire the design, research, and development of social interfaces that make LLMs easy and enjoyable to use. 
\begin{acks}
  This work was supported in part by an Apple Scholars in AI/ML PhD fellowship and J.P. Morgan PhD Fellowship.
  We are extremely grateful to Kaan Sancak, Jaemarie Solyst, Xinzhi Jiang for stress-testing our tool.
  We appreciate Tianqi Chen, Charlie Ruan, and Alex Cabrera for their suggestions and support for integrating on-device LLMs.
  We express our gratitude to Alex Bäuerle, Upol Ehsan, Muhammed Fatih Balin, Luis Morales-Navarro, Wesley Hanwen Deng, Young Wu, Giorgio Pini, Justin Blalock, Viraj Kacker, Seongmin Lee, Ben Hoover, Anthony Peng, Alec Helbling, Matthew Hull, Mansi Phute, Harsha Karanth, Pratham Mehta, Joanna Cheng, Lena Do, Smera Bhatia, Angelina Wang and Parul Pandey for their advocacy and valuable feedback.
\end{acks}

\balance
\bibliographystyle{ACM-Reference-Format}
\bibliography{24-wordflow}

\end{document}